\documentclass[12pt]{article}
\usepackage{graphicx}
\usepackage{color}
\newcommand{\beq}{\begin{equation}}
\newcommand{\eeq}{\end{equation}}
\newcommand{\bea}{\begin{eqnarray}}
\newcommand{\eea}{\end{eqnarray}}

\newcommand{\gsim}{\lower.7ex\hbox{$
\;\stackrel{\textstyle>}{\sim}\;$}}
\newcommand{\lsim}{\lower.7ex\hbox{$
\;\stackrel{\textstyle<}{\sim}\;$}}

\newcommand{\eod}{\end{document}}

\definecolor{verm}{rgb}{0.8,0.1,0.0}

\begin{document}
\thispagestyle{empty}
\vspace*{-22mm}

\begin{flushright}

UND-HEP-13-BIG\hspace*{.08em}02\\

\end{flushright}

\vspace*{1.3mm}

\begin{center}
{\Large {\bf Memorial Contribution about Kolya Uraltsev's Talk at 2012 Workshop $B \to D^{**}$}}

\vspace*{19mm}

{\bf I.I.~Bigi$^a$} \\
\vspace{7mm}
{\sl Notre Dame, IN 46556, USA}

\vspace*{-.8mm}

{\sl email address: ibigi@nd.edu}

\vspace*{10mm}

{\bf Abstract}\vspace*{-1.5mm}\\
\end{center}
Uraltev has given us a deeper understanding of the forces that produce  the data from beauty and charm hadrons for the last 35 years and neutron EDM in the future -- in particular about non-perturbative QCD.  At the Workshop in the November 2012 before his death he had focused mostly 
on the data about semi-leptonic $B$ decays beyond $B \to l \nu D/D^*$. 
Here is a very short  review of his talk at this Workshop. Next year colleagues will produce a `Memorial Book' about Kolya Uraltsev's excellent contributions to understand fundamental dynamics.

\vspace{3mm}

\hrule

\tableofcontents
\vspace{5mm}

\hrule\vspace{5mm}

\section{Introdution}

The Workshop $B \to D^{**}$ in the Nomember 2012 was the last conference that Kolya Uraltsev has attended \cite{KOLYATALK}. My plan was to attend also this WS -- mostly to discuss fundamental physics with Kolya in person; however I was not able to do it. I was connected with Kolya by internet even close to his death.  

While I have worked mostly about inclusive decays of beauty and charm hadrons, but 
obviously I was following Kolya's work about exclusive ones (even now in my future) and discussed it with him in person and by internet. 
He had used very powerful, refined and {\em new} theoretical tools for inclusive and exclusive ones. 
He was a real pioneer -- and still is -- about heavy quark transitions:  heavy quark symmetry and heavy quark expansions (HQE), sum rules including those from small velocity (SV) and 
spin rules etc. etc. etc.

There are several central items discussed at this workshop:  
\begin{itemize}
\item  
The issues are the impact of broad resonances and continua final states about the rates of 
inclusive and exclusive semi-leptonic $B$ decays.  Is there a tension between `measured' 
$|V_{cb}|_{\rm incl}$ and $|V_{cb}|_{\rm excl}$? 

\item 
Can we probe $B \to \tau \nu D/D^*$ for a sign of New Dynamics (ND) now and for the future? 
\item
The impacts of dynamics affect non-leptonic $B$ decays more indirectly and subtly, but strongly.

\item 
There are several parameters that are smaller than unity, namely $\bar \Lambda/m_c$, 
$\bar \Lambda/m_b$ (with $\bar \Lambda = M_B - m_b$), SV used by HQET, 
BPS and of course $\alpha_S$ depending on the scales. Therefore one can use expansions; it depends on the situations which of those give the leading source of dynamics.  

\item 
A more general point: how much do we understand the inner structures of strong forces in a 
quantitative way?

\item 
For a practical way: how well can we use correlations of the lengths of the sides of `the' 
CKM triangle and its angles with precision?

\end{itemize}
There is another issue for me, namely: Kolya Uraltsev has not been given even close to 
fair credit for his works and impact. History might change that.

\section{Excited Heavy-Quark Mesons}

Operator Product Expansion (OPE) and in particular HQE are the best 
tools to deal with non-perturbative dynamics in inclusive rates with matrix elements of local operators 
based on real QCD, not just models. In HQE several parameters $\bar \Lambda$, $\mu^2_{\pi}$, 
$\rho_D^3$ ... are predicted (or at least limited) by one number from the data, namely 
hyperfine splitting $M_{B^*} - M_B \simeq 47$ MeV. It is `nice' that the data show us 
$M_{B^*}^2 - M_B^2 \sim M_{D^*}^2 - M_D^2$ 
\footnote{Data give further:  
$M_{B^*}^2 - M_B^2 \sim M_{D^*}^2 - M_D^2 \sim M_{K^*}^2 - M_K^2 
\sim M_{\rho}^2 - M_{\pi}^2$ -- a miracle?}. 

These HQE parameters are `dynamics-driven' terms that 
include kinetic ones -- i.e., phase spaces. 
This point was discussed again recently with details \cite{SLW}.

For a long time it was said to compare inclusive rates for $B \to l \nu X_c$ 
vs. the sum of exclusive ones in 
semi-leptonic $B$ decays. The sum of $B \to l \nu D/D^*$ and the {\em narrow} resonances 
$B \to l \nu D^{**}$ are close to the inclusice ones -- but not enough. 
One has to deal with broad resonances and continuum final states. Also one has to calculate form factors in $B \to D/D^*$ more precisely; the correlations with broad resonances and continuum ones  depend in their definitions; furthermore their differences are not clear with 
finite data and experimental uncertainties. 

Uraltsev had dealt with non-perturbative QCD for a long time and also had given 
reference to our colleagues in Orsay \cite{YAOU,JUG,YAOU2} at the workshop in November 2012. 
From the beginning of the 21th century Uraltsev had worked on the impact of $P-waves$ to close the gap using several theoretical tools, not just models \cite{KOLYA1,KOLYA2}. In a paper published in 2007 \cite{PENE} it was compared 
predictions vs. data. Some experts  
said that the impact of $P-waves$ should be stronger for $j^P = 3/2^+$ amplitudes than 
for $j^P = 1/2^+$ ones. However the data seen to disagree, and even fans of symmetry 
(like Kolya and I) say that in the end real data are the best judges 
\footnote{As in the world highest courts need long times to produce good decisions.}. 

At the 2012 Workshop $B \to D^{**}$ Uraltsev had focused on the impact of {\em radial} excitations 
($j^P = 1/2^+$) and $D-waves$ states ($j^P = 3/2^+$) using {\em non-local correlators} in $B$ mesons. 
He showed the enhanced inclusive yields of radial and $D-waves$ in the final states of 
$b \to c l \nu$ decays. It might resolve the $1/2$ vs. $3/2$ `paradox' and produces correlations of the formfactors in $B \to D^*$ and $B \to D$ transitions. 
Two papers show how to understand the theoretical tools that had to be used there \cite{GMU1,GMU2}. 
They discuss SV heavy quark sum rules including spin rules for relativistic bound-states in details 
(and deal with non-local correlators).  
In particular the second paper is not short at all; honesty tells you that only really deep `students' should read it.  

We cannot say that these problems had already been solved -- however his talk shows us the 
direction where we need more data, more work and accuracies. `En route' he had shown that several 
technologies help us  for the future in practical ways and 
also give a deeper understanding the features of QCD. 

\section{Inclusive vs. exclusive rates \& CP/T asymmetries}
\label{CONN}

Obviously the 2012 Workshop $B \to D^{**}$ focuses on exclusive semi- and non-leptonic $B$ transitions. It also gives important lessons about QCD dynamics in general. On the other hand 
it is part of a larger landscape in fundamental physics: 
\begin{itemize}
\item 
Now Kolya said there is no real tension $|V_{cb}|_{\rm incl}$ and 
$|V_{cb}|_{\rm excl}$ using new technologies. We have to think about it in the future.
\item
If there is a real tension between $|V_{cb}|_{\rm incl}$ and $|V_{cb}|_{\rm excl}$, one has to worry even 
more for $|V_{ub}|_{\rm incl}$ and $|V_{ub}|_{\rm excl}$. It would show that our control of 
non-perturbative QCD is more limited as thought. 
\item
There is a very practical reason: the values of $V_{ub}$, $V_{cb}$ and their ratio extracted from inclusive 
and exclusive ones are crucial to construct `the' CKM triangle (out of six ones from six quarks) from SM. 
That has obvious impact on measured CP asymmetries in $B$ decays whether SM can produce it alone or not. Furthermore the situation is subtle for CP violation. We know that ND can at best produce non-leading source of CP violation in beauty decays. The impact of ND might 
hide in the `CKM' triangles which we construct.

Therefore we need the best available tools or produced new theoretical technologies. Kolya has worked about it as a pioneer for twenty years. 

\item
Kolya had worked about $SU(3)$ symmetry and breaking in beauty and charm hadrons 
including non-perturbative effects in a theory, not just using diagrams \cite{SLW}. 

\item
Kolya had thought and gave papers about the neutron EDM starting in 1985 \cite{ANS}, 
worked with me in 1990 \cite{1990EDM} and with Th.Mannel in 2011/2012 \cite{2012MANNEL}.

\end{itemize}

These comments are not given here to brag what Kolya did, but to show his broad  
horizon for more thirty years. The last item about EDM shows that he does not give up; 
he tells us we need more precision for good reasons.

\section{Summary}

The neutral Higgs state have been found now in the region predicted by the SM. 
However we still have the same reasons to find ND somewhere and some scales due to 
neutrino oscillations, matter vs. antimatter asymmetry, dark matter and dark energy. Therefore we have to measure transitions 
with more energy, more jets \& their topologies, but also more accuracies in direct and indirect ways -- like correlations 
about rare decays and CP asymmetries and probe the `CKM' {\em triangles} with more precision. 
We know that the SM gives at least the leading source of known matter. 
For that we need more and better theoretical tools and check them with local operators 
and non-local  correlations. Kolya did it in more than thirty years in deep ways.  
It is not always easy to read his papers -- but that price comes with a real {\em prize}!

Finally a lot of work is still needed -- and a lot of thinking!

\vspace{0.5cm}

{\bf Acknowledgments:} This work was supported by the NSF under the grant number 
NSF PHY-1215979.

\vspace{4mm}


\end{document}